\documentstyle[pre,aps]{revtex}        
\input epsf
\tighten
\begin{document}

\twocolumn[\hsize\textwidth\columnwidth\hsize\csname @twocolumnfalse\endcsname
\title{Correlation functions in the two-dimensional\\  random-field Ising model}
\author{
S. L. A. de Queiroz$^a$\footnote{e-mail:
sldq@if.uff.br. Address from August 1999: Instituto de F\'\i sica,
Universidade Federal do Rio de Janeiro, Caixa Postal 68528, 21945-970 
Rio de Janeiro RJ, Brazil; e-mail:sldq@if.ufrj.br} and
R. B. Stinchcombe$^b$\footnote{e-mail: stinch@thphys.ox.ac.uk}
 }
\address{
$^a$ Instituto de F\'\i sica, Universidade Federal Fluminense,\\ Avenida
Litor\^anea s/n, Campus da Praia Vermelha, 24210-340 Niter\'oi RJ, Brazil \\
$^b$ Department of Physics, Theoretical Physics, University of Oxford,\\
 1 Keble Road, Oxford OX1 3NP, United Kingdom
}
\date{\today}
\maketitle
\begin{abstract}
Transfer-matrix methods are used to study the probability
distributions  of spin-spin
correlation functions $G$ in the two-dimensional random-field Ising
model, on long strips of width $L = 3 - 15$ sites, for binary  field
distributions at generic distance $R$, temperature $T$ and field
intensity $h_0$.
For moderately high $T$, and $h_0$
of the order of magnitude used in most experiments, the
distributions are singly-peaked, though rather asymmetric.
For low temperatures the single-peaked shape deteriorates,
crossing over towards a double-$\delta$
ground-state structure. 
A connection is obtained between the probability distribution
for correlation functions and the underlying distribution of  
accumulated field fluctuations. Analytical expressions are in good
agreement with numerical results for $R/L \gtrsim 1$, low $T$,  
$h_0$ not too small, and near  $G=1$.
From a finite-size {\it ansatz}
at $T=T_c (h_0=0)$, $h_0 \to 0$,  averaged 
correlation functions are predicted to scale with 
$L^y h_0$, $y =7/8$. From numerical data we estimate
$y=0.875 \pm 0.025$,
in excellent agreement with theory.
In the same region, the RMS relative width $W$
of the probability distributions varies for fixed $R/L=1$
as $W \sim h_0^{\kappa}\, f(L\, h_0^u)$ with $\kappa \simeq
0.45$, $u \simeq 0.8$ ; $f(x)$ appears to  
saturate when $x \to \infty$, thus implying $W \sim
h_0^{\kappa}$ in  $d=2$.
\end{abstract}

\twocolumn
\narrowtext
\vskip0.5pc]

\section{Introduction}
\label{intro}

It is by now well established that the space dimensionality $d=2$ is the
lower critical dimension of the random field Ising model 
(RFIM)~\cite{imb84,bri87,aiz89}, in agreement with the early domain-wall picture
of Imry and Ma~\cite{imr75}. Thus, as usual for a borderline
dimensionality, details of two-dimensional behaviour are
rather intricate. The divergence of the
low-temperature correlation length as the field intensity approaches zero is 
apparently anomalously severe~\cite{aha83}. This is at least partly
responsible for difficulties encountered in the application of  
normally very powerful numerical techniques to this problem. In
particular, transfer-matrix (TM) methods have been 
used, either for fully finite~\cite{mbh81,pyt85,fer85} or 
semi-infinite~\cite{gla86} geometries. TM calculations have usually focused upon
the structure factor, as obtained from suitable derivatives of the partition
function. The correlation length is then derived from the structure factor,
under assumption of specific scaling forms~\cite{pyt85,fer85,gla86}; results
thus far have been at least in qualitative agreement with theoretical 
predictions~\cite{aha83}.

Many recent studies of the RFIM, both in $d=2$ and 3, have concentrated on 
zero-temperature  properties, as an exact ground-state
algorithm first applied some time ago~\cite{ogi86} has been
revisited~\cite{zerot1,zerot2}. In our earlier work~\cite{us1,us2},
where a domain-wall scaling picture was developed for bar-like systems
in general $d$, numerical support for theory was provided
in $d=2$, $T=0$ by a version of such algorithm adapted to strip geometries. 
For $T \neq 0$ we relied on a TM treatment of the free energy, again on
strips.   

Here we deal directly with probability distributions of
spin-spin correlation functions, calculated by TM
methods on semi-infinite (strip) systems. 
Interest in probability distribution functions has increased recently,
regarding extensive quantities in critical disordered systems. This is
in line with the growing realisation that lack of self-averaging tends
to be the rule, rather than the exception, e.g. for susceptibilities
and magnetisations in such 
systems~\cite{nsa}, implying that
the width of the associated probability distributions is
a permanent feature that does not trivially vanish with increasing
sample size. 
In the present case, lack of self-averaging does not come as a surprise, 
as correlation functions are not
extensive~\cite{derrida}, so the usual Brout argument~\cite{brout}
is not expected to apply.
Also, in $d=2$ the random field moves the second-order transition to
$T=0$, so the $d=2$ RFIM is off criticality at any $T \geq 0$;
experimental manifestations of microscopic features of the $d=2$ RFIM
come indirectly through (sample-averaged)
non-critical properties~\cite{bir83,fer83,bel85}.
Indeed, consideration of the crossover behaviour in the vicinity of
the zero-field, pure-Ising, critical point provides interesting
information, as shown in Section~\ref{tc0}~.  

In what follows, we first discuss the ranges of spin-spin distance $R$, 
temperature $T$ and random-field intensity $h_0$ for which the
statistics of correlation functions display the most interesting features,
and  illustrate our choices with simple examples. We then turn to the 
connection between field-- and correlation-function distributions, and
show how, in suitable limits, one can extract the latter from the former.
Next we study the line $T =T_c(h_0=0), h_0 \to 0$, and use correlation
functions to extract information on scaling behaviour corresponding
to the destruction of long-range order by the field.
A final section summarizes our work. 

\section{Numerical techniques and parameter ranges}
\label{numtech}

We calculate the spin-spin correlation function 
$G(R) \equiv \langle \sigma_0^1 \sigma_R^1 \rangle$, between spins on the same
row (say, row 1), and $R$ columns apart, of strips of a square lattice
of ferromagnetic Ising spins with nearest-neighbour interaction $J=1$,
of width $3 \leq L \leq 15$
sites with periodic boundary conditions across.
This is done along the lines of Sec. 1.4 of 
Ref.~\onlinecite{nig90}, with standard adaptations for an inhomogeneous
system~\cite{sldq}. The strip widths used are those manageable on standard 
workstations, without unusually large memory or time requirements; 
as the main overall advantage of TM calculations  
(against e.g. those on fully finite, $L \times L$ systems), 
is that monotonic trends set in for relatively small strip widths~\cite{nig90},
the upper bound on $L$ does not significantly constrain our analysis. It
does, however, matter for the values of $R$ used, since the interesting
range of $R/L$ is around one, where the transition between $d=1$
and $d=2$ behaviour takes place. 
 
Different sorts of averaging are involved in this case.
For a given realisation of the site-dependent random fields,
one has  (for sufficiently low field intensity)
a macroscopic ground-state degeneracy~\cite{zerot1}. TM
methods take into account  the Boltzmann weights of all possible
spin configurations,
so  they scan the whole set of available ground states for a given
realisation of quenched disorder. One must then promediate over many
such realisations, which is done as follows. At each iteration of the TM
from one column to the next, the random-field values $h$  are drawn for each 
site from the binary distribution:
\begin{equation} 
P(h) = {1 \over 2}[\, \delta (h - h_0) + \delta (h + h_0)\, ]\ \  .
\label{eq:pd}
\end{equation}
By shifting the origin along the strip and accumulating the respective
results, one can produce normalized histograms, $P(G)$, 
of occurrence of $G(R)$.
With typical strip lengths $N=10^6$ columns, we generate $10^4 - 10^5$
independent estimates of $G(R)$ for $R$ in the range $5 - 15$ which
corresponds to $R/L \sim 1$, as explained above.
 
In our previous study of the
unfrustrated random-bond Ising model~\cite{dqrbs}, the probability
distribution function of correlations was expected to be log-normal
for strictly one-dimensional systems~\cite{derrida}. This led
us to a picture where, for strip width $L$ and spin-spin
distance $R$, the distribution would evolve perturbatively 
away from log-normal with increasing $L/R$. Thus, there we used
logarithmic binning for the histograms of occurrence of $G(R)$: a 
convenient interval of variation of $\ln G(R)$ was divided into, usually,
$10^3$  bins, each particular realization being assigned to the
appropriate bin.  As a similar starting point is not available here,
and negative values of correlations may occur, we have
resorted to a simple linear choice, dividing the whole $[ -1, 1]$ interval of
variation of $G(R)$ into (again, usually $10^3$) equal bins. 

The temperature and field intervals of interest are broadly circumscribed
because spin-spin correlations are induced by the ferromagnetic (unit)
interaction. Thus one must keep to values of $T$ and $h_0$ that are not
sufficient to render the coupling negligible; rough boundaries, to be
refined next, are $T_c(h_0=0)= 2.269 \dots$ and $h_{0c}(T=0)=4$ 
(above this latter value each spin always obeys the local field).

We have found $T_{low} = 0.6$ to be low enough
to display ground-state effects rather prominently.
 Recall that strictly {\it at} $T=0$ 
correlation-function histograms are trivial double--$\delta$ peaks at 
$G(R) =\pm 1$; this reflects the frozen-domain structure
of the ground state, which is best
investigated directly as done by others~\cite{zerot1,zerot2}.
Conversely, here we wished to investigate departures
from the double-$\delta$ shape, induced by increasing $T$.
On the other hand, $T_{high} = 2.0$  is high enough so that field
fluctuations (in the range of $h_0$ spelt out in the next paragraph)
have mainly a perturbative effect. 
\begin{figure}
\epsfxsize=8,4cm
\begin{center}
\leavevmode
\epsffile{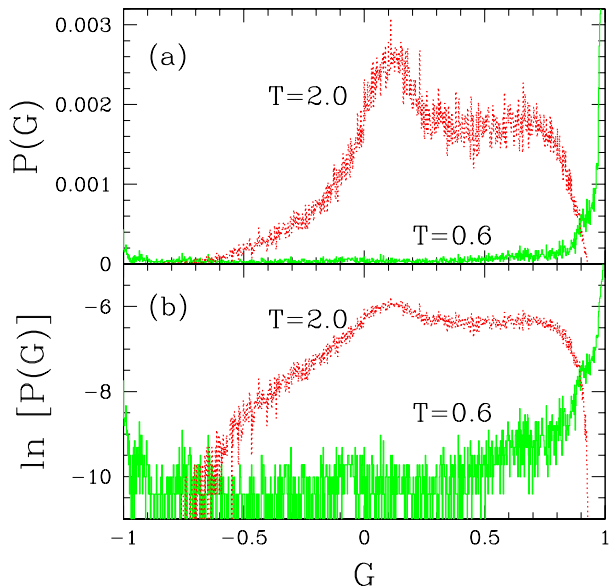}
\caption{
Normalized histograms $P(G)$ of occurrence of $G$ 
for strip width $L=5$,
length $N=10^6$ columns, $R=15$, $h_0=0.5$, and $T= 0.6$ and $2.0$.
Binwidth $2 \times 10^{-3}$.
Vertical axis has linear scale
in (a) and logarithmic in (b), the latter
in order to emphasize values occurring with low frequencies. 
}
\label{fig:hist}
\end{center}
\end{figure}
Experimental studies~\cite{bir83,fer83,bel85} of $d=2$ dilute Ising 
antiferromagnets in a uniform field $H$ (argued  by
Fishman and Aharony~\cite{fis79} to be equivalent to the RFIM) concentrate on
$H$-values corresponding to $h_0 \lesssim 0.1 - 0.2$ in Eq.~(\ref{eq:pd}),
enough to cause significant departures from zero-field behaviour.
Higher fields $h_0 \gtrsim 1$ are convenient to enrich domain 
statistics in simulations of fully finite systems, as they
reduce low-temperature domain sizes and increase
degeneracy~\cite{zerot1,zerot2}. 
However, already for
$h_0 = 0.5$  the histograms of correlation functions were found to
be utterly distorted (compared to a paradigm of
single-peaked structures with reasonably-defined widths), so as to be
intractable in terms of a simple description with few parameters.
This echoes the experimental observation for Rb$_2$C$_{0.7}$Mg$_{0.3}$F$_4$,
that ``...applied fields very much less than the Co$^{2+}$ molecular
fields ... have quite drastic effects''~\cite{bir83}. 
Fig.~\ref{fig:hist} illustrates the point. 

\begin{figure}
\epsfxsize=8,4cm
\begin{center}
\leavevmode
\epsffile{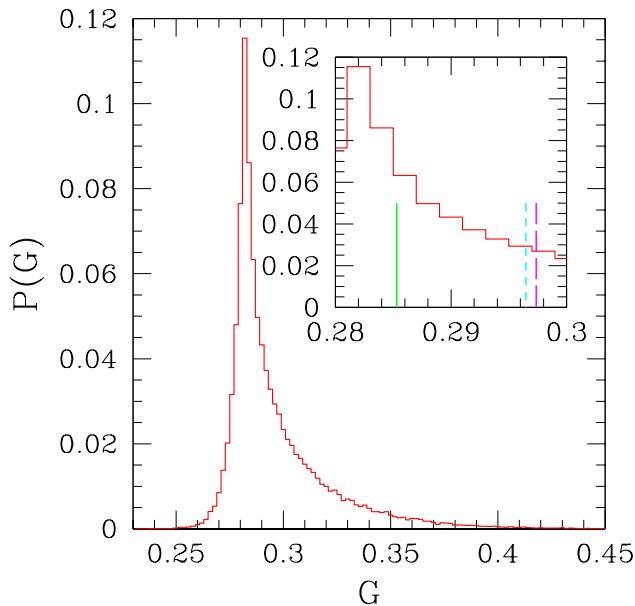}
\caption{
Normalized histogram $P(G)$ of occurrence of $G$ for strip width $L=5$,
length $N=10^6$ columns, $R=15$, $h_0=0.05$ and $T=2.0$.
Binwidth $2 \times 10^{-3}$.
Vertical bars in inset located respectively 
at: $G_0$ (full line); $\exp \langle \ln G \rangle$ (short-dashed); 
$\langle G \rangle$ (long-dashed).}
\label{fig:hight}
\end{center}
\end{figure}

Nevertheless, for  $h_0 \lesssim 0.1 - 0.15$ and high $T$, the
overall picture stays very close to that depicted in
Fig.~\ref{fig:hight},
with the following main features: (i) a clearly-identifiable single peak,
{\em below} the zero-field value $G_0 \equiv G(h_0=0)$; (ii) a short
tail below the peak and a long one above it, such  that
(iii) all moments of order $\geq 0$ of the distribution are {\em above}
$G_0$. In Figure~\ref{fig:hight} we show the zeroth ($\exp \langle \ln G
\rangle$) and first ($\langle G \rangle$) moments.
This scenario breaks down for low temperatures, 
as $G(h_0=0)$ becomes close to the upper limit of unity, for $R/L \sim 1$
and the strip widths within reach. However, this latter regime can
be understood in terms of a direct connection between field-- and 
correlation-function distributions, described in Section~\ref{cfdvshd}
below.

\begin{figure}
\epsfxsize=8,4cm
\begin{center}
\leavevmode
\epsffile{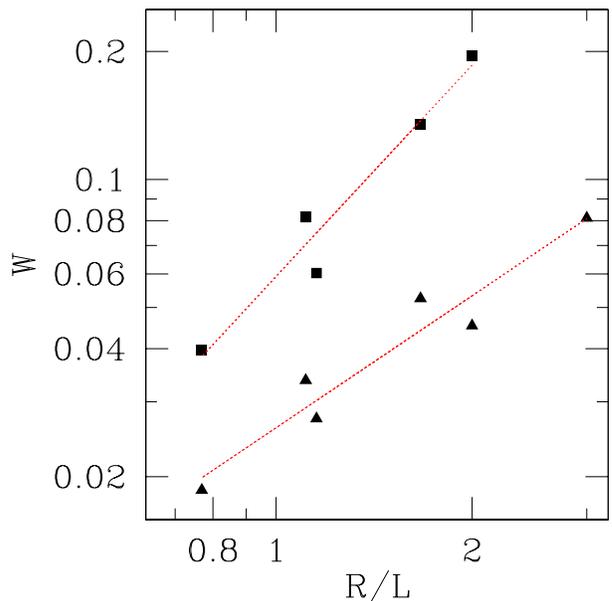}
\caption{
Double-logarithmic plot of RMS relative widths $W$ of distributions
against $R/L$. Strip widths $L=5$, 9, 13; $R=10$, 15;
$h_0=0.05$ (triangles) and 0.15 (squares). Straight lines are
least-squares fits to each set of data (see text). $T=2.0$. 
}
\label{fig:wvslr}
\end{center}
\end{figure}

For fixed $R$, small $h_0$ and high $T$, Figure~\ref{fig:narrow} shows
the typical evolution of distributions against $L$.  Note that, 
with $h_0=0.15$, the
single-peak structure shows early signs of fraying. 
\begin{figure}
\epsfxsize=8,4cm
\begin{center}
\leavevmode
\epsffile{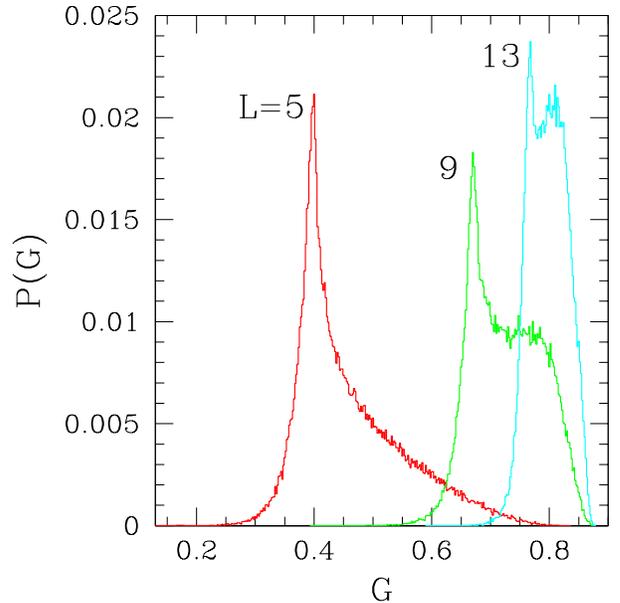}
\caption{
Normalized histograms $P(G)$ of occurrence of $G$ for strip widths $L=5$,
$9$ and $13$; length $N=10^6$ columns, $R=10$, $h_0=0.15$ and $T=2.0$. 
Binwidth $2 \times 10^{-3}$.
}
\label{fig:narrow}
\end{center}
\end{figure}
One can see in Figure~\ref{fig:narrow} a
narrowing effect with increasing $L$. This is quantitatively depicted
in Figure~\ref{fig:wvslr}, for which the use of $R/L$ on the horizontal
axis is inspired in usual ideas of finite-size scaling, 
and has proven fruitful in our earlier study of random-bond
systems~\cite{dqrbs}. 

Though the RMS relative width 
$W\equiv(\langle G^2 \rangle - \langle
G \rangle^2)^{1/2}/\langle G \rangle$
appears to approach zero for small $R/L$ (where, as argued in
Ref.~\onlinecite{dqrbs}, $d=2$ behaviour should show up), we note
that: (i) the data display 
non-monotonic jumps even for fixed $h_0$; and (ii) power-law fits of
$W$ against $(R/L)^x$ show a rather strong dependence of $x$ on $h_0$
(for data in Figure~\ref{fig:wvslr} one has $x \simeq 1$ for $h_0=0.05$,
and $2$ for $h_0=0.15)$.
These facts indicate that, from consideration of the above data alone, we
are not in a position to conclude that $W \to 0$ as the true $d=2$
regime is approached. Indeed, in Section~\ref{tc0} below a different
analysis, at fixed $R/L$, strongly suggests that the widths do not vanish
in the two-dimensional limit.    

\section{Distribution of $G$ from field distribution}
\label{cfdvshd}

An important question is how the field distribution gives rise to the
distribution for the correlation function $G(R) \equiv \langle \sigma_0
\sigma_R \rangle$ (at specific separation $R$).

A scenario worth exploring is the following: the probability distribution
$P(G)$ for $G(R)$ arises from a distribution of characteristic scales
$\xi$, related to $G(R)$ via $G(R) \sim \exp(-R/\xi)$, with $\xi$
distributed according to some distribution. This last probability
distribution has then to be related to the field distribution. At low
temperatures a domain picture might provide that relationship: a
distribution of domain sizes $\xi_i$ arises from the distribution of
fields aggregated over each domain, {\it e.g.} by minimising energy
(or free energy) along the lines of Ref.~\onlinecite{us1}, but
generalised to consider specific field configurations, with their
associated probability (the free energy minimisation may make such
an approach applicable up to temperatures of order $T_c(h=0)$\ ).

The simplest such scheme uses a common domain size $\xi$, over which
the total field is $h = x h_0 \sqrt{\xi L}$ with 
$p(x) = e^{-x^2/2}/\sqrt{2\pi}$ .
This is the distribution of aggregated fields on a domain, arising
from the independent distributions 
${1 \over 2}[ \delta (h - h_0) + \delta (h + h_0) ]$ of fields $h_i$
at each site $i$..
Then minimising the free energy per unit length (for the $T=0$
problem) gives
\begin{equation}
\xi = \xi(x) = {4J^2L \over x^2h_0^2}\ \ \ .
\label{eq:xi0}
\end{equation}
\noindent Hence, from the probability distribution $p(x)$, there arises
a probability distribution for $\xi(x)$, and via that a probability
distribution  $P(G)$ for $G(R) \sim \exp(-R/\xi)$. The result is
\begin{equation}
P(G,T=0) = \left( { 2J^2L \over h_0^2R}\right)^{1/2} {1 \over
\sqrt{2\pi}}
{G^{{ 2J^2L \over h_0^2R}-1} \over (\ln{1/G})^{1/2}}\ \ \ .
\label{eq:pG0}
\end{equation}
The important parameter in this zero-temperature description is
$\lambda \equiv { 2J^2L \over h_0^2R}$ (which is
of order one for $R=15$, $L=5$, $h_0 = 0.5$, for example).

The $T \neq 0$ generalisation of such pictures involves the entropic
contribution $-T(S_0+S_1)$ to the free energy $F$,
which includes a contribution
from positioning
of domain walls ($-TS_0$) (see Ref.~\onlinecite{us1}, but still allowing
for probabilities of specific field configurations) and also one from the
random-walk-like wandering of the domain walls ($-TS_1$).

These entropies are (using the simplest picture of a single $\xi(x)$)
$S_0 = k_B \ln\xi(x)/\xi(x)$ (using reduction valid for $\xi(x)$
large), and $S_1 = k_B \ln\mu^L/\xi(x) = k_B (L/\xi(x))\ln\mu$,
with $\mu \sim z-1$, $z=$ lattice coordination number.
Minimisation of $F$ per unit length then gives  
\begin{equation}
0 = {h_0 x \sqrt{L\xi} \over 2JL} -1  +  {k_BT \over JL}\left( L \ln\mu +
\ln\xi -1\right)\ \ \ .
\label{eq:min}
\end{equation}
\noindent The variable $x$ is again distributed with 
the domain-aggregated field distribution $p(x) = e^{-x^2/2}/\sqrt{2\pi}$ 
which, via Equation~(\ref{eq:min})
then provides the distribution of $\xi$ and finally the distribution
for $G \sim e^{-R/\xi}$ (along the general lines indicated above).
Different pairs of terms dominate Equation~(\ref{eq:min}) in different
regimes of $h_0$, $T$ and $L$. Of special interest to us are the
first-order low-temperature corrections. An approximate treatment
of Equation (\ref{eq:min}), valid for $G$ near 1, gives
\begin{equation}
P(G,T) \propto P(G,T=0) (\ln 1/G)^{-4k_BT/JL}\ \ \ ,
\label{eq:pGT}
\end{equation}
\noindent with $P(G,T=0)$  given  by Equation (\ref{eq:pG0}).
Apart from weakly $L$-dependent normalization factors, one should have
\begin{equation}
P^{\prime}(G) \equiv P(G,T) G^{-\alpha}(\ln(1/G))^{\beta}= {\rm const}\ \
\ ,
\label{eq:constant}
\end{equation}
\noindent where $\alpha \equiv (2J^2L/h_0^2R)-1$,
$\beta \equiv 1/2 + 4k_BT/JL$\ .
\begin{figure}
\epsfxsize=8,4cm
\begin{center}
\leavevmode
\epsffile{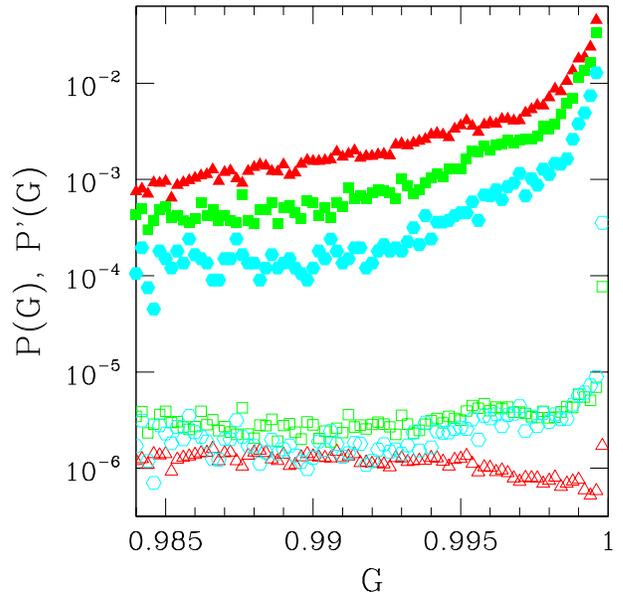}
\caption{Histograms of occurrence of $G$ for $T=0.6$, $h_0=0.5$, $R=15$,
near $G=1$; $L=3$ (triangles), $5$ (squares) and $7$ (hexagons).
Binwidth $2 \times 10^{-4}$. Full symbols: normalized histograms, $P(G)$.
Empty symbols: $P^{\prime}(G)$, Eq.~(\protect{\ref{eq:constant}}).
}
\label{fig:cfdvsh}
\end{center}
\end{figure}
In Figure~\ref{fig:cfdvsh} we check Equation~(\ref{eq:constant})
 for $T=0.6$, $h_0=0.5$, $R=15$ and $L=3$, 5 and 7.  Use of narrow
strips (i.e. $R/L >1$) and high fields is important in order to produce
broad distributions in the low-temperature regime concerned.  One sees
that, indeed, the strong $G$-- dependence of  $P(G,T)$ near $G=1$ can be 
essentially accounted for by the factors in 
Equation~(\ref{eq:constant}).

\section{Scaling near zero-field critical point}
\label{tc0}

According to theory~\cite{aha83,fer83,bel85,fis79,aha78}, the
scaling behaviour
of the RFIM depends on the variable $h_0^2 |t|^{-\phi}$ where
$h_0$ is the random-field intensity and
$t =(T-T_c(h_0))/T_c(h_0)$ is a reduced temperature. For
 $d > d_c =2$ $T_c(h_0)$ is the field-dependent temperature at which 
a sharp transition still occurs; it turns out that
even in $d=2$ the dominant terms still depend on the same
combination, where now~\cite{bel85} ``$T_c(h_0)$''
denotes a pseudo-critical temperature marking, e.g., the location of
the rounded  specific-heat peak. This is true except for the $d=2$
specific heat (which does not concern us directly here), where 
$\ln h_0$-dependent terms also play an important role~\cite{fer83,nvl76}.
Further, it is predicted~\cite{aha78} that the crossover exponent
$\phi = \gamma$, the pure Ising susceptibility exponent. In $d=2$,
specific heat~\cite{fer83} and neutron-scattering~\cite{bel85} data 
are in good agreement both with the choice of scaling variable as
above, and with the exactly known $\gamma =7/4$ .

\begin{figure}
\epsfxsize=8,4cm
\begin{center}
\leavevmode
\epsffile{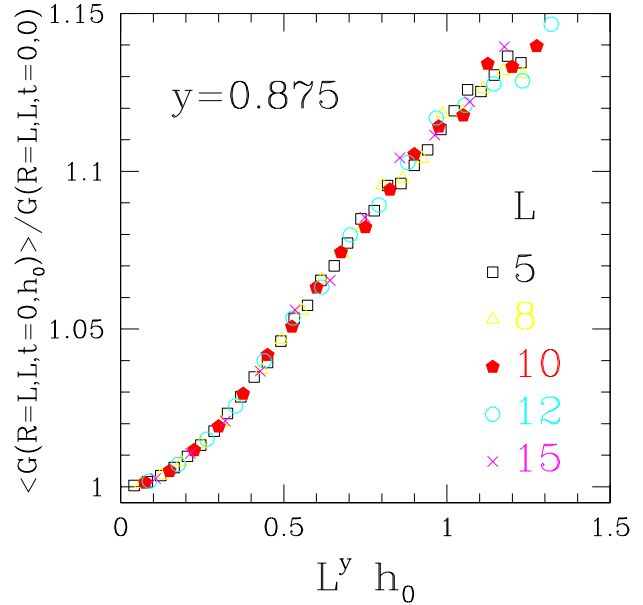}
\caption{Averaged correlation functions (normalised by their zero-field
counterparts) against $L^y\, h_0$ with $y=0.875$. Each point is the
central estimate on a strip $N=10^5$ sites long. See text and Table I
for a discussion of estimated error bars.
}
\label{fig:tccfa}
\end{center}
\end{figure}

Here we propose a direct check of scaling, as follows. For $h_0 \to 0$,
near $T_{c\, 0}\equiv T_c(h_0=0)$, one expects~\cite{fer83} 
``$T_c(h_0)$''$= T_{c\, 0} - c h_0^{2/\phi}$. Hence,
$ h_0^{-2/\phi} t \simeq  h_0^{-2/\phi}(T - T_{c\,0})$ apart from a
small, finite shift.
Setting $T = T_{c\, 0}$ and making the usual 
finite-size scaling {\it ansatz}~\cite{bar83} $t \to L^{-1/\nu}$
with the pure Ising  value $\nu = 1$ (this latter assumption to be 
verified), one obtains that the (finite-size) scaling
variable at $T=T_{c\, 0}$ must be
\begin{equation}
x \equiv h_0 L^{\phi/2\nu}\ , \ \ \ (T=T_{c\, 0}, h_0 \to 0)
\label{eq:scalevar}
\end{equation}
\noindent with $\phi/2\nu= 7/8$ in $d=2$. This implies that the
correlation length related to the
decay of ferromagnetic spin-spin correlations diverges along this 
particular line as
\begin{equation}
\xi(T=T_{c\, 0}, h_0 \to 0) \sim h_0^{-1/y}\ ,\ \ \   y=\phi/2\nu.
\label{eq:xih}
\end{equation} 
From standard finite-size scaling~\cite{bar83}, the correlation functions
for distance $R$, strip width $L$, $t \equiv T - T_{c\,0} =0$
and random-field intensity $h_0$ are then expected to behave as
\begin{equation}
G(R,L, t=0,h_0)   = L^{-\eta}\, \Gamma (R/L,L^y h_0)\ \ \ . 
\label{eq:gscale}
\end{equation}
In Figure~\ref{fig:tccfa} we show, for fixed $R/L=1$, the scaling plot
thus
suggested, where $y$ has been adjusted to provide
the best data collapse. The same procedures have been used very
recently in studies of unfrustrated random-bond Potts
models~\cite{oy99}.
Note the use of averaged correlation
functions, $\langle G \rangle$. We also performed plots with 
{\it typical} ones~\cite{dqrbs}, $\exp(\langle \ln G \rangle)$, with 
entirely similar results. As remarked in Section~\ref{numtech}, one
has $\langle G \rangle > G(h_0=0)$, on account of the long forward
tail of the distribution. This happens for $\exp(\langle \ln G \rangle)$
as well, and is a scenario valid only for low field intensities. Near
the end of the region where scaling holds, on the right 
of Figure~\ref{fig:tccfa}, one indeed sees the beginning of a trend  
towards stabilization (which would, for higher fields, presumably
turn into a decreasing function of $h_0$, were scaling still valid). 

The value $y =0.875$ used in Figure~\ref{fig:tccfa} gave the
best data collapse, which still kept reasonably good over the
interval ($0.85,0.90$). The plots using $\exp(\langle \ln G \rangle)$
behaved in the same way. Thus our estimate is: $y = 0.875 \pm 0.025$~,
in very good agreement with the finite-size scaling {\it ansatz}
described above, with $\gamma=7/4$, $\nu=1$. 
\begin{table}
\caption{
Correlation functions calculated at $T=T_{c\,0}$, distance $R=L$
and random-field intensity $h_0=0$ ($G_0$) and $h_0 = 0.8L^{-7/8}$
($\langle G(h_0) \rangle$). Error bars in parentheses give
uncertainties in last quoted digits, from spread among central
estimates for five different runs on strips with $N=10^5$.}
\vskip 0.2cm 
 \halign to \hsize{
\hfil#\quad\hfil&\quad\hfil#\quad\hfil&\hfil#\quad\hfil&\hfil#\quad\hfil\cr
    $L$ & $G_0$ & $\langle G(h_0)\rangle$ & $\langle G(h_0)\rangle/G_0$
 \cr\noalign{\smallskip}
   5 & $0.333422277$ & $0.36420(31)$& $1.0923(10)$\cr
   8 & $0.300005458$ & $0.32800(30)$& $1.0933(10)$\cr
  10 & $0.284437852$ & $0.31086(40)$& $1.0929(14)$\cr
  12 & $0.272124932$ & $0.29733(48)$& $1.0926(18)$\cr
  15 & $0.257635774$ & $0.28148(59)$& $1.0926(23)$\cr
}
\end{table}

Each point in Figure~\ref{fig:tccfa} represents an average taken
from one run on strips $N = 10^5$ columns long. We now discuss the
estimation of error bars, not shown in the Figure.
Recalling that the
width of the distributions is not expected to vanish in the
thermodynamic limit, we follow the lines extensively elaborated
elsewhere for similar cases~\cite{sldq,dqrbs,sbl}, and estimate
fluctuations
by evaluating the spread among overall averages (i.e. central estimates)
from different samples. For values of $L$ and $h_0$
such that $L^{7/8} h_0 = 0.8$ (approximately midway along
the horizontal axis of Figure~\ref{fig:tccfa}), we performed
series of five runs, each with $N=10^5$,  for each $L$. Table I
shows the results. 
One sees that Equation~(\ref{eq:gscale}) is satisfied to
within 2 parts in $10^3$. Such an agreement is further evidence in suport
of the scaling {\it ansatz} proposed above;
it also  suggests that the scaling power is  $y=7/8$ exactly.

Incidentally, note that from the constancy against $L$ of the ratio
$\langle G(R,L,t=0,h_0)\rangle/G(R,L,t=0,0)$, as verified in 
Table I, and the scaling of correlation functions given in
Equation~(\ref{eq:gscale}), one immediately has $\eta=
\eta_{Ising}=1/4$
for the decay of ferromagnetic correlations at $T=T_{c\,0}$, 
$h_0 \to 0$~.

We now return to scaling of the RMS relative width $W$  of the
distribution against field and strip width, restricting ourselves
to $T$ near $T_{c\, 0}$ and $h_0$ not very large. 
For fixed $R/L$, taking into account that
the distribution broadens (a) with increasing random-field
intensity (which is elementarily expected), and (b) also with increasing
strip width (which we noticed in our numerics at $T=T_{c\, 0}$),
we propose the following scaling form:
\begin{equation}
W = h_0^{\kappa}\, f(L\, h_0^u)\ \ \ ,
\label{eq:wscale} 
\end{equation}
\noindent where the effective length $L_h \equiv h_0^{-u}$ plays the role
of a saturation distance, such that $f(x) \to {\rm constant}, x \gg 1$.
In other words, (i) for high temperatures such as $T=T_{c\, 0}$
and small $h_0$ there must be a regime in which the distribution
remains recognizably similar to Figure~\ref{fig:hight}, with the
field-induced broadening reaching a relatively small maximum as 
$R$,$L \gg h_0^{-u}$ (at fixed $R/L$). At the other end $x \ll 1$,
the only obvious constraint is that (ii) $f(x)$ must not increase 
faster than $x^{-\kappa/u}$ as $x \to 0$, if it does diverge at all.

\begin{figure}
\epsfxsize=8,4cm
\begin{center}
\leavevmode
\epsffile{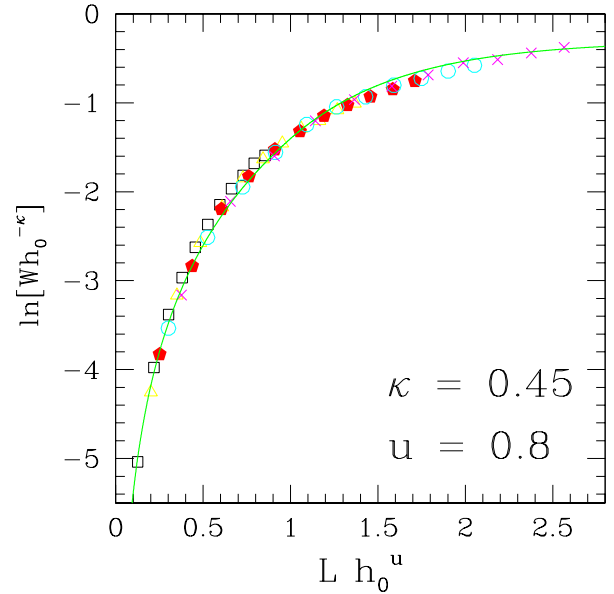}
\caption{Semi-logarithmic scaling plot of RMS relative widths, 
$W\, h_0^{\kappa}$
against $L\, h_0^u$.  Key to symbols is the same as in 
Figure~\protect{\ref{fig:tccfa}}. Curves are
fitting splines (see text).
}
\label{fig:wvslh}
\end{center}
\end{figure}

From scaling plots of $W\, h_0^{-\kappa}$ against $L\, h_0^u$
(at $T = T_{c\, 0}$ and $h_0$ not very large)
with tentative values of the exponents,
we have found the best data collapse to occur for $\kappa \simeq
0.43 - 0.50$ and $u \simeq 0.8$. Figure~\ref{fig:wvslh}, where the
vertical
axis is logarithmic, shows our
results for $\kappa=0.45$ and $u=0.8$. For $x >1$, the fitting spline is
the function
$y = -0.3 -5.3\exp(-1.57x)$, implying a limiting scaled width 
$W\, h_0^{-\kappa} = \exp(-0.3) = 0.83$~, consistent with (i) above.
For $x <1$ the fitting curve is given by $y = 1.73\ln x - 1.40$, in
agreement with requisite (ii). 

To our knowledge there is no structural relationship between the
width exponents $\kappa$ and $u$, and the standard critical
indices, such as the crossover exponent $\phi$
discussed above.
 Conversely, one would expect widths to 
behave similarly to the above picture even at $T \neq T_{c\, 0}$,
provided that one keeps to high temperatures and low field intensities.
Most likely, asymptotic scaled widths will depend on $T$; 
a matter for further investigation is
whether or not the numerical values of the exponents will also vary.

\section{conclusions}
\label{conc}

We have  studied the probability distributions of occurrence of spin-spin
correlation functions $G$ in the $d=2$ RFIM, for binary distributions of
the local fields, at generic distance $R$, temperature $T$ and field
intensity $h_0$, on long strips of width $L = 3 - 15$ sites.

We have shown that for moderately high temperatures, of the order of
the zero-field transition point $T_{c\,0}$, and field intensities 
$h_0 \lesssim 0.1-0.2$ in units of the nearest-neighbour
coupling (the same order of magnitude used in most experiments), the
distributions retain a recognizable single-peaked structure, with
a well-defined width. However, they display considerable asymmetry,
with a short tail below the maximum and a long one above it, the latter
owing to the mutual reinforcement between ferromagnetic spin-spin
interactions and large accumulated-field fluctuations. For low
temperatures the single-peaked shape deteriorates markedly, as
crossover takes place towards the double-$\delta$ structure 
characteristic of the ground state.

We have established a connection between the probability distribution
for correlation functions and the underlying distribution of  
accumulated field fluctuations. Starting  from a zero-temperature
description based on the distribution of (essentially flat) domain
walls across the strip, we have shown how (low-) temperature effects
can be incorporated, and proposed analytical expressions for the
main dependence of the distribution of correlation functions on
$R$, $L$, $T$ and $h_0$. In their assumed domain of validity, i.e. 
$R/L \gtrsim 1$, $T \ll 1$,  not very small $h_0$, and close to
the upper extreme $G=1$, they are  in good
quantitative agreement with numerically calculated  distributions.

At $T=T_{c\,0}$, for $h_0 \to 0$, we have made contact with
scaling theory for bulk systems, and developed a finite-size
{\it ansatz} to describe the scaling behaviour of averaged 
correlation functions. The variable that describes such behaviour 
was found to be $L^y h_0$, with $y=0.875 \pm 0.025$ from numerical data,
in excellent agreement with the {\it ansatz}'s prediction, $y =7/8$.
In the same region, we have also studied the RMS relative width $W$
of the probability distributions, and found that, for fixed $R/L=1$
it varies as $W \sim h_0^{\kappa}\, f(L\, h_0^u)$ with $\kappa \simeq
0.45$, $u \simeq 0.8$ . We have shown that $f(x)$  fits well to
a saturating form when $x \to \infty$, thus implying $W \sim
h_0^{\kappa}$ in  $d=2$.
 
Further developments of the present work would include: (i)
establishing analytical expressions to connect field fluctuations,
domain size distribution and correlation function distributions in 
regimes such as $R/L \simeq 1$ (relevant to $d=2$ behaviour), $T \sim 
T_{c\,0}$, and valid for generic $G$; and (ii) a systematic study of
the variation of widths and their associated exponents, both against
temperature and the ratio $R/L$. We are currently considering such
extensions. 

Finally, as regards contact with experiment, one may ask how the present
results for correlation functions relate e.g. to the wavevector-dependent
scattering amplitudes in neutron scattering~\cite{bir83}. Attempts in
this direction have been made earlier~\cite{gla86}. Since the
scattering function reflects spatial averages
over relatively extended regions, a connection to correlation functions 
must be established  via a correlation
length which represents the average decay of spin-spin
correlations~\cite{gla86,bir83}. Furthermore, fitting
numerical data from one end to experimental results from the other is a
tricky task, which is usually mediated by resorting to heuristically
proposed line shapes. Of these, Lorentzian and Lorentzian-squared
functions have been  among the most popular~\cite{gla86,bir83}, though
in principle there is no reason  why one must
be restricted to them. A broad range of possible line
shapes, compounded with the wide variation exhibited by several
properties of correlation functions, as shown in the present work,
causes one to anticipate a fairly involved investigation.
\acknowledgements

SLAdQ thanks the Department of Theoretical Physics
at Oxford, where this  work was initiated, for the hospitality, and
the cooperation agreement between Conselho Nacional
de Desenvolvimento Cient\'\i fico e Tecnol\'ogico and
the Royal Society for funding his visit. 
Research of SLAdQ
is partially supported by the Brazilian agencies Minist\'erio da Ci\^encia
e Tecnologia, Conselho Nacional
de Desenvolvimento Cient\'\i fico e Tecnol\'ogico and Coordena\c c\~ao de
Aperfei\c coamento de Pessoal de Ensino Superior.

RBS thanks Instituto de F\'\i sica, UFF for warm hospitality, and the 
Royal Society for support, during a visit to further this research.
Partial support from EPSRC Oxford Condensed Matter Theory Rolling Grant
GR/K97783 is also acknowledged.

\end{document}